\documentclass{aa}  

\usepackage{graphicx}
\usepackage{graphics}
\usepackage{txfonts}
\bibpunct{(}{)}{;}{a}{}{,} 
\usepackage{natbib}
\usepackage{enumerate}
\bibpunct{(}{)}{;}{a}{}{,} 
\begin{document}

   \title{A gas-rich AGN near the centre of a galaxy cluster at $z \sim 1.4$
\thanks{Based on observations carried out with the IRAM Plateau de Bure Interferometer. 
IRAM is supported by INSU/CNRS (France), MPG (Germany), and IGN (Spain).}}


\author{V. Casasola\inst{1},
          L. Magrini\inst{2},
          F. Combes\inst{3},
          A. Mignano\inst{1},    
           E. Sani\inst{2},    
	  R. Paladino\inst{4,1},
           \and
        F. Fontani\inst{2}
          }

   \institute{INAF -- Istituto di Radioastronomia \& Italian ALMA Regional Centre, Via P. Gobetti 101, 40129 Bologna, Italy \\
              \email{casasola@ira.inaf.it}
         \and
              INAF -- Osservatorio Astrofisico di Arcetri, Largo E. Fermi 5, 50125 Firenze, Italy
              \and
              Observatoire de Paris, LERMA, 61 Av. de l'Observatoire, 75014, Paris, France
              \and
              Dipartimento di Fisica e Astronomia, Universit\`{a} di Bologna, Viale Berti Pichat 6/2, 40127, Bologna, Italy
              }
   \date{Received; accepted }

\titlerunning{AGN.1317}
\authorrunning{Casasola et al.}

 
\abstract
{The formation of the first virialized structures in overdensities dates back to $\sim$9~Gyr ago, i.e. in the redshift range $z \sim 1.4 - 1.6$.  
Some models of structure formation predict that the star formation activity in clusters was high at that epoch, implying large reservoirs of cold molecular gas.}
{Aiming at finding a trace of this expected high molecular gas content in primeval clusters, 
we searched for the $^{12}$CO(2--1) line emission in the most luminous active galactic nucleus (AGN) of the cluster 
around the radio galaxy 7C 1756+6520 at $z \sim 1.4$, one of the farthest spectroscopic confirmed clusters. 
This AGN, called AGN.1317, is located in the neighbourhood of the central radio galaxy at a projected distance of $\sim$780~kpc.}
{The IRAM Plateau de Bure Interferometer was used to investigate the molecular gas quantity in  AGN.1317,  observing 
the $^{12}$CO(2--1) emission line.}
{We detect CO emission in an AGN belonging to a galaxy cluster at $z \sim 1.4$. 
We measured a molecular gas mass of 1.1~$\times$10$^{10}$~M$_{\odot}$, comparable to that found in submillimeter galaxies.
In optical images, AGN.1317 does not seem to be part of a galaxy interaction or merger.
We also derived the nearly instantaneous star formation rate (SFR) from H$\alpha$ flux obtaining a SFR~$\sim$65~M$_{\odot}$~yr$^{-1}$. 
This suggests that AGN.1317 is actively forming stars and 
will exhaust its reservoir of cold gas in $\sim$0.2-1.0~Gyr.}
{}


   \maketitle
%

\section{Introduction \label{sec:intro}}

Galaxy clusters are the largest collapsed structures in the Universe
with total masses up to 10$^{15}$~M$_{\odot}$ \citep[e.g.][]{arnaud09}.
One of the most important issues related to the study of galaxy clusters is
the evolution of their star formation (SF) activity.
Some models of galaxy evolution in dense environments predict that the SF in clusters
was high at the epoch of the formation of galaxy clusters, i.e. $z \sim 1.5$, and later at $z \sim 1$
it was rapidly quenched \citep[e.g.][]{martig08,taranu12,wetzel13}.
One key requirement for SF is the presence of a reservoir
of dense, cold gas that can be efficiently converted into stars.   
This is especially crucial for galaxies in rich clusters, because they are expected 
to be affected by mechanisms able to remove cold gas from the haloes and discs 
of infalling galaxies \citep[e.g. ram pressure stripping,][]{gunn72} or to prevent further 
cooling of gas within galaxies' dark matter haloes 
\citep[starvation or strangulation, e.g.][]{larson80,bekki02}.
This environmental dependence profoundly influences the evolutionary 
histories of galaxy clusters and the star formation--galaxy
density relation represents  a well-established observational hallmark of how galaxies
evolve as a function of environment.

\begin{figure*}
\centering
\includegraphics[width=0.8\textwidth]{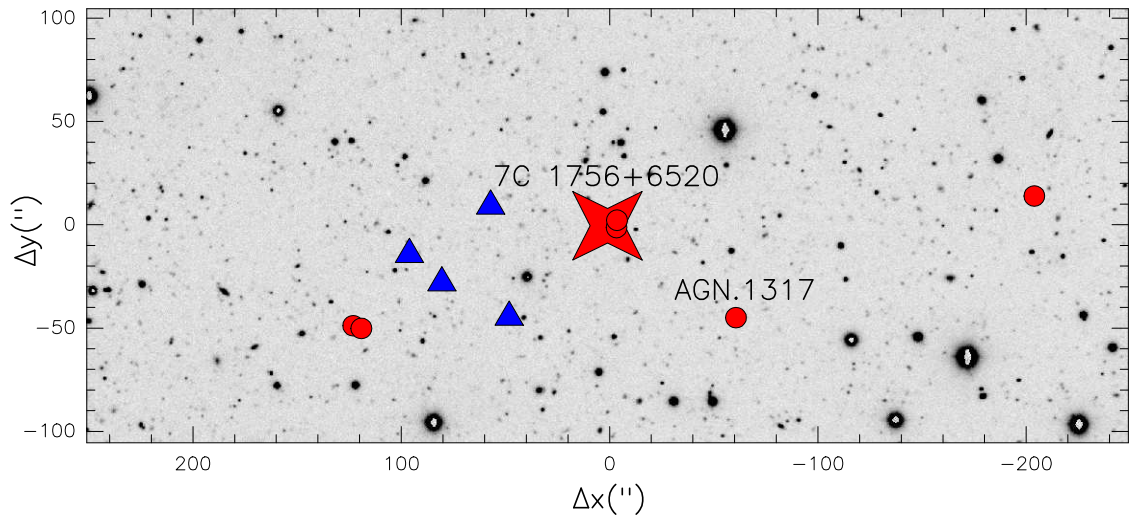}
\caption{Location of the galaxies belonging to Group1 ($z \sim$ 1.42) and Group2 ($z \sim$ 1.44) in the $B$-band image (NOAO).
Galaxies of Group1 are represented by red circles, while those of Group2 by blue triangles. 
The central radio galaxy 7C 1756+6520 is the red star and belongs to Group1.
North is at the top and east to the left.
The field of view is $\sim$500$^{\prime\prime}$ $\times$ 200$^{\prime\prime}$ centred on the radio galaxy.
1$^{\prime\prime}$ corresponds to $\sim$8.4~kpc.
}
\label{fig:image-cluster}
\end{figure*}

Studies based on multi-wavelength tracers of SF activity have shown that SF 
decreases with increasing galaxy density at $z < 1$ 
\citep[e.g.][]{hashimoto98,ellingson01,gomez03,patel09}.
At intermediate redshift ($z \sim 0.2$) luminous infrared galaxies are preferentially
forming stars at the outskirts of some massive clusters \citep[e.g.][]{haines10}, 
suggesting that SF may be quenched within the central regions at these epochs.   
As we approach the epoch when galaxies should be forming the bulk of their stars 
($z > 1$), there is some controversy in literature on the star formation--density relation for clusters at $z > 1$. 
Some works found that the above relation 
reverses at $z > 1$ \citep{elbaz07,cooper08,hilton10,tran10}, while others that the period between $1 < z < 2$ 
is the one when active SF is suppressed in the very central regions of galaxy clusters \citep{strazzullo10,bauer11,grutzbauch12}.
In the first scenario, the redshift range $z \sim 1.4 - 1.6$ seems to correspond to the epoch in which galaxy clusters showed 
the highest star formation rate (SFR), especially in their central regions, in contrast to what is 
observed at $z < 1$, while in the second scenario star-forming galaxies are seen at the outskirts of the clusters.
Furthermore, the number of cluster galaxies with one or more AGN has
been found to increase out to redshifts $z > 1$, and they lie preferentially near the cluster centre 
\citep[e.g.][]{galametz09,martini09,martini13}.
The SF and AGN activity in high-$z$ cluster galaxies should, therefore, both be fueled by significant reservoirs of molecular gas.
Observations of this gas in high-$z$ clusters can provide insights into the processes governing galaxy evolution
in dense environments.

So far,  molecular gas has been detected in distant ($z > 1$) galaxies in a hundred objects  
\citep[see the review by][]{solomon05}, but only few of them belong to clusters.
The most distant detection is related to a  $z = 7.08$ quasar host galaxy, where the fine structure line of  
{\sc [C~ii]} 158~$\mu$m and the thermal dust continuum were detected by \citet{venemans12}.  
For cluster galaxies the situation is the following: at intermediate redshift, \citet{geach09,geach11} 
presented $^{12}$CO(1--0) detections in five starburst 
galaxies in the outskirts of the rich cluster Cl~0024+16 at $z \sim 0.4$, and \citet{jablonka13} 
reported CO detections in three luminous infrared galaxies belonging to the clusters CL~1416+4446 at $z \sim 0.4$ 
and CL~0926+1242 at $z \sim 0.5$. 
At higher redshift, \citet{wagg12a} presented $^{12}$CO(2--1) missing detections toward 
two dust-obscured galaxies with AGN and a serendipitous 
detection in an obscured AGN belonging to a cluster at $z \sim 1$.
Very recently, \citet{emonts13} found a $^{12}$CO(1--0) detection in the so-called Spiderweb Galaxy
at $z \sim 2$, one of the most massive systems in the early Universe and surrounded by a dense web  
of proto-cluster galaxies.
The Ly$\alpha$ nebulae, called Ly$\alpha$ blobs (LABs) at $z \sim 2-6$, have also been the object of molecular gas observational campaigns, 
 collecting for the moment non-detections or tentative 
detections \citep[e.g.][]{chapman04,matsuda07,tamura09,yang12,wagg12}. 
Their nature is at present still poorly understood, but they could represent   proto-clusters hosting an AGN at their centre \citep[e.g.][]{cen12}. 
Finally, using ALMA \citet{wagg12b} detected the {\sc [C~ii]} line and thermal dust emission 
from a pair of gas-rich galaxies at $z = 4.7$, BR1202-0725, 
a possible proto-cluster, but detected only dust continuum from a third companion whose redshift remains 
unknown.
To summarise, the only secure CO detection in a high-$z$ ($\sim$1) cluster AGN is, to our knowledge, that given in  \citet{wagg12a}.

In this work, we present the first observations of the $^{12}$CO(2--1) emission line toward an AGN, AGN.1317, 
belonging to a galaxy cluster at $z \sim 1.4$, obtained using the IRAM Plateau de Bure Interferometer (PdBI). 
This paper is structured as follows.
In Sect. \ref{sec:sou}, we present the main properties of AGN.1317.
In Sect. \ref{sec:obs}, we describe our CO observations, and the corresponding results
are collected in Sect. \ref{sec:res}.
The discussion of results is presented in Sect.~\ref{sec:dis}, and finally
we summarise this work in Sect.~\ref{sec:con}.   
In the present paper, we adopt the following values for the cosmological constants, $H_{0} = 70$ ~km~s$^{-1}$~Mpc$^{-1}$,
$\Omega_{\rm m} = 0.3$, and $\Omega_{\Lambda} = 0.7$ corresponding to the $\Lambda$CDM cosmology. 
With these values, 1$^{\prime\prime}$ corresponds to $\sim$8.4 kpc at $z \sim 1.4$.

\begin{table*}
\caption[]{Fundamental characteristics of the overdensity  of galaxies  associated with 7C 1756+6520 and of AGN.1317.}
\begin{center}
\begin{tabular}{lccccc}
\hline
\hline
Source       & $\alpha_{\rm J2000}$     & $\delta_{\rm J2000}$ & $z({\rm G10})$  & $z({\rm M12})$   \\
\hline
7C 1756+6520 & 17$^{\rm h}$57$^{\rm m}$05.44$^{\rm s}$ & 65$^{\circ}$19$^{\prime}$53\farcs11 & 1.4156 & -\\
Group1           &                                         &                                     & $\sim$1.42 & - \\
Group2           &					       &                                     & $\sim$1.44 & - \\	
\hline
AGN.1317     & 17$^{\rm h}$56$^{\rm m}$55.75$^{\rm s}$ & 65$^{\circ}$19$^{\prime}$07\farcs00 & 1.4162 $\pm$ 0.0005 & 1.4168 $\pm$ 0.0005  \\							
\hline
\hline
\end{tabular}
\label{table1}
\end{center}
\tablefoot{($\alpha_{\rm J2000}$, $\delta_{\rm J2000}$) are from G10 and those of
AGN.1317 coincide with the phase tracking centre of our $^{12}$CO(2--1) observations.
$z({\rm G10})$ and $z({\rm M12})$ indicate the redshift determinations from G10 and M12, respectively.}
\end{table*}

\section{The source: AGN.1317 \label{sec:sou}}
\citet{galametz09,galametz10} have isolated and spectroscopically confirmed,   
through the optical Keck/DEep Imaging Multi-Object Spectrograph (DEIMOS),
an overdensity  of galaxies associated with the radio galaxy 
7C 1756+6520 at $z = 1.4156$.
In addition to the central radio galaxy, they confirmed twenty-one galaxies 
with spectroscopic redshifts consistent with that of 7C~1756+6520. 
In the field around the radio galaxy the velocity dispersion is rather large, up to $\sim$13000~km~s$^{-1}$ 
\citep[][hereafter G10]{galametz10}, and the ensemble has not yet relaxed into one big structure.   
For this reason, the overall structure is better defined as an overdensity, while   
 two distinct smaller substructures have been identified and designated as clusters of galaxies:
one of seven galaxies (Group1) with redshift similar to that of the central radio galaxy
($z \sim 1.42$), and a more compact one at $z \sim 1.44$ (Group2).
Galaxies belonging both to Group1 and Group2 are within 2~Mpc from the radio galaxy, 
while most of other galaxies belonging to the large-scale overdensity are more than 2~Mpc away from
the radio galaxy.
Figure \ref{fig:image-cluster} shows a portion of the galaxy overdensity associated with 7C 1756+6520.
In this figure, cluster galaxies belonging to Group1 and Group2 are clearly marked with different symbols and colours.

Six of the  spectroscopically confirmed galaxies of the overdensity, including the central 
radio galaxy, are AGN, and AGN.1317 is one of them.
\citet{magrini12} (hereafter M12)  have presented near infrared (NIR) spectroscopic observations of AGN.1317, 
obtained with the Large Binocular Telescope (LBT).
The galaxy AGN.1317 belongs to Group1 having a redshift $z = 1.4162$ (G10, see Table~\ref{table1})  and it is located in the neighbourhood of 
the central radio galaxy with a projected distance from it of  $\sim$780~kpc (see Fig.~\ref{fig:image-cluster}).
Among the photometrically studied AGN, AGN.1317 is the brightest one (B = 20.16, G10).
Its NIR spectrum shows clear broad features, such as H$\alpha$ and H$\beta$, associated with 
the broad-line region, and several forbidden lines, such as the {\sc [O~iii]}, {\sc [N~ii]}, 
and {\sc [S~ii]} doublets, associated with the AGN narrow-line region and/or with the host star-forming galaxy.    
Moreover, the {\sc [O~iii]} lines show a clear asymmetric profile with prominent blue-shifted wings, 
signature of a strong gas outflow (${\rm v} > 1000$~km~s$^{-1}$) possibly driven 
by the AGN radiation pressure. 
The prominent NIR emission of AGN.1317 (F$_{\rm H{\alpha}}$(narrow) = 2.8 $\times$10$^{-16}$~erg~cm$^{-2}$~s and
F$_{\rm H{\alpha}}$(broad) = 9.7$\times$10$^{-16}$~erg~cm$^{-2}$~s, M12)
makes this galaxy an excellent candidate for CO detection in a galaxy cluster at such high redshift.

The galaxy overdensity around 7C 1756+6520  was observed  in radio-cm continuum 
(e.g, 74~MHz: Cohen et al.~2007; 1.4~GHz: Condon et al.~1998; 
4.8~GHz: Becker et al.~1991), but the low spatial resolution of these observations ($\sim$45--80$^{\prime\prime}$) 
did not allow AGN.1317 to be resolved. On the other hand, the only high spatial resolution (1\farcs5) observations did not reach 
the sensitivity needed to detect AGN.1317 \citep[4.8~GHz:][]{lacy92}.
The main characteristics of AGN.1317 and of the overdensity  of galaxies  associated with 7C 1756+6520
are collected in Table~\ref{table1}.

\section{Observations \label{sec:obs}}
We observed AGN.1317 
with the IRAM PdBI (six antennas) in the compact C configuration of the array on November 22--23, 2012.
We tuned the receivers to 95.4135~GHz ($\sim$3~mm), the frequency of the
$^{12}$CO(2--1) emission line (230.538~GHz) at redshift $z = 1.4162$.
The spectral setup used during observations provides a velocity coverage of $\sim$3000~km~s$^{-1}$
(i.e. $\sim$1~GHz).
A total time of 8 hours were spent on-source and the observing conditions were excellent 
in terms of atmospheric phase stability (PWV $<$ 2~mm).
Typical system temperatures were around 75--110~K during the observations.
The radio source 2200+420 was used for bandpass calibration, 1849+670 for flux calibration, and 1716+686 
for phase and amplitude calibrations. 

The data were reduced and analysed with the IRAM/GILDAS\footnote{http://www.iram.fr/IRAMFR/GILDAS/} 
software \citep{guilloteau00}.  
Data cubes with 256 $\times$ 256 pixels (0\farcs2 pixel$^{-1}$) were created over the velocity interval 
of $\sim$3000~km~s$^{-1}$ in bins of $\sim$60~km~s$^{-1}$. 
The half power primary beam (field of view) is $\sim$53$^{\prime\prime}$ at the observed
frequency.
The image presented here was reconstructed with natural weighting and restored with a Gaussian beam of 
dimensions 4\farcs73~${\times}$~2\farcs33 ($\sim$40~kpc $\times$ 20~kpc) and PA = 147$^{\circ}$.
In the cleaned map, the rms level is 0.27\,mJy\,beam$^{-1}$ at a velocity resolution of 
$\sim$60\,km\,s$^{-1}$.
At a level of 3$\sigma$, no 3~mm continuum was detected toward AGN.1317 down to an rms noise level of
0.04~mJy~beam$^{-1}$.
The conversion factor between intensity and brightness temperature is $11\,{\rm K\,(Jy\,beam^{-1})^{-1}}$ at 95.4135~GHz.
Since AGN.1317 is at the centre of the image, the map presented here is not corrected 
for primary beam attenuation.

\begin{figure*}
\centering
\includegraphics[width=0.4\textwidth]{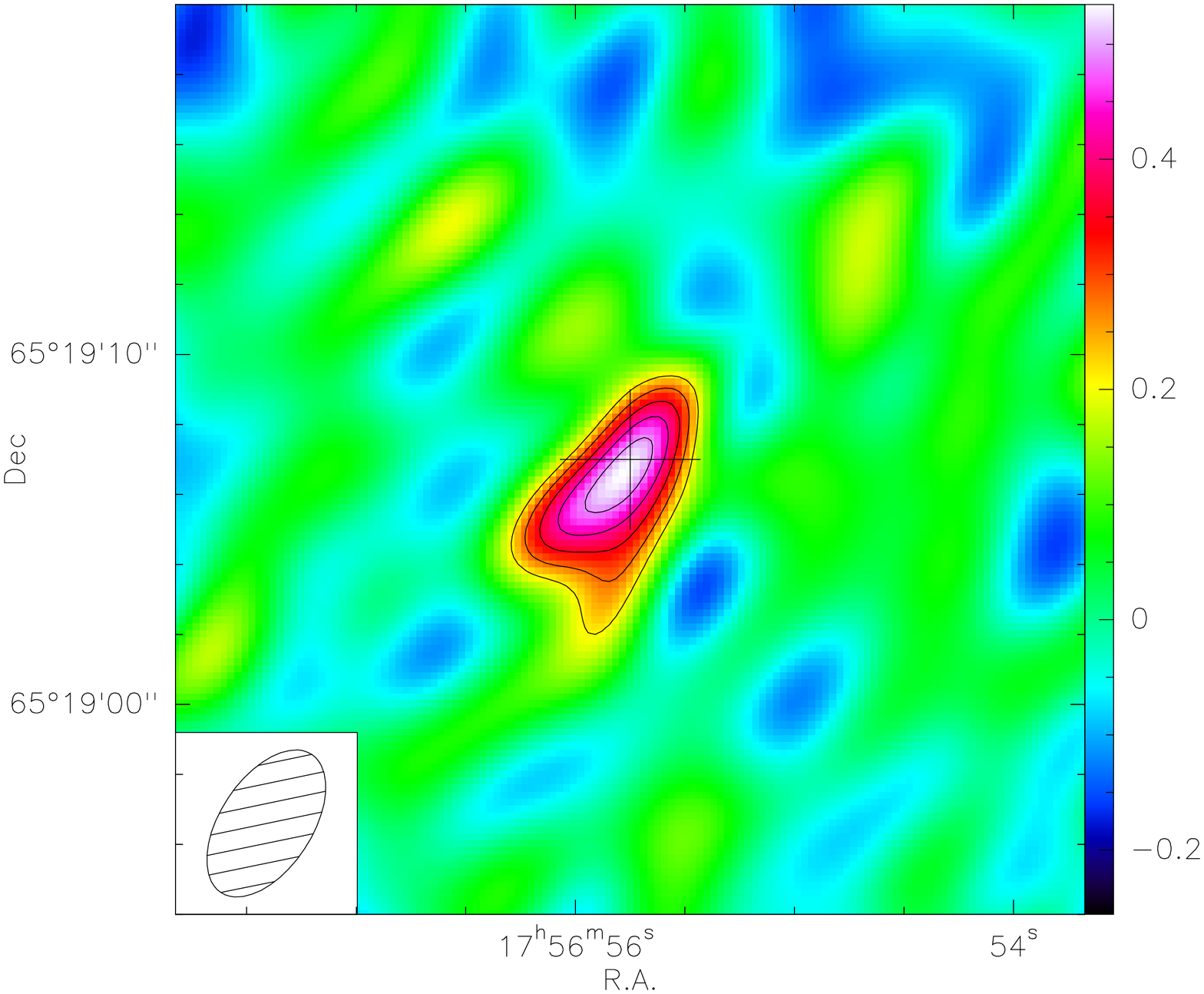}
\hspace*{1cm}
\includegraphics[width=0.4\textwidth]{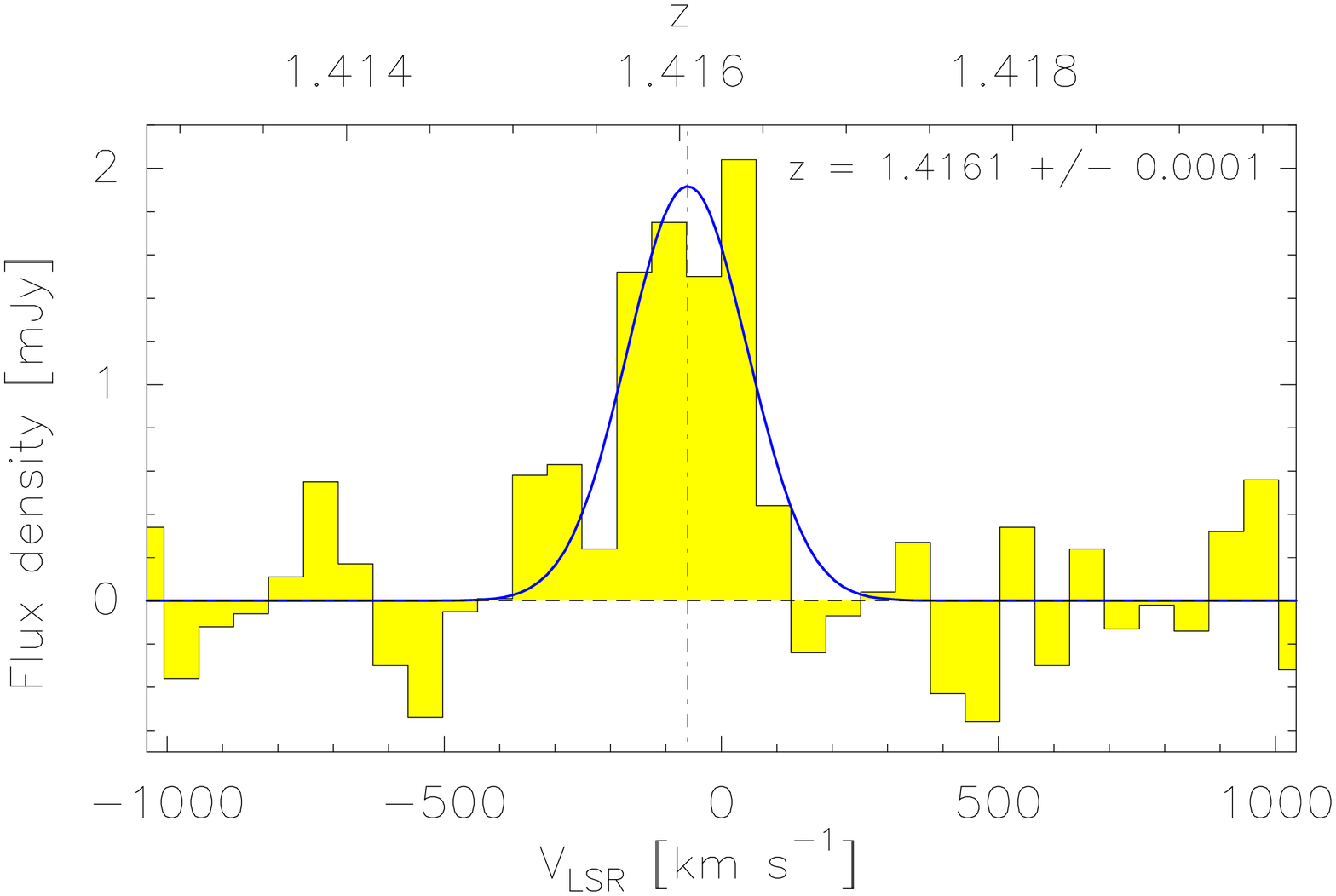}
\caption{\textit{Left Panel:} $^{12}$CO(2--1) intensity map obtained with the IRAM PdBI toward AGN.1317. 
The color wedge of the intensity map is in Jy~beam$^{-1}$~km\,s$^{-1}$.
The black cross marks the coordinates of the phase tracking centre of our observations (see Table~\ref{table1}).
The {\it rms} noise level is $\sigma = 0.07\,{\rm Jy\,beam^{-1}\,km\,s^{-1}}$ 
and contour levels run from 3$\sigma$ to 7$\sigma$ with 1$\sigma$ spacing.
In this map a velocity range of $\sim$360~km~s$^{-1}$ is used.
The beam of 4\farcs73~${\times}$~2\farcs33 (PA = 147$^{\circ}$)
is plotted at lower left.
\textit{Right Panel:} $^{12}$CO(2--1) integrated spectrum obtained with the IRAM PdBI toward AGN.1317
centred on the optical velocity of AGN.1317, $ {\rm V_{opt}} = 4.25 \times 10^{5}$~km~s$^{-1}$ ($z = 1.4162$, G10).  
The spectrum has a flux density (S$_{\rm CO}$) scale from $-0.7$ to 2.2~mJy, 
a velocity scale from $-1050$ to 1050~km~s$^{-1}$ (lower x-axis), and a $z$ scale from 1.4128 to 1.4197 
(upper x-axis).
The Gaussian fit and the vertical dot-dashed line show that the $^{12}$CO(2--1) emission is 
blue-shifted by $\sim$61~km~s$^{-1}$ with respect to the optical emission (G10).}
\label{fig:map}
\end{figure*}

\begin{table*}
\caption[]{Results from IRAM PdBI observations for AGN.1317.}
\begin{center}
\begin{tabular}{ccccccc}
\hline
Line  & Position\,\tablefootmark{a} & $z$ (This work)         &  $\Delta$${\rm v}_{\rm FWHM}$ \tablefootmark{b}   & S$_{\rm CO}$$\Delta$${\rm v}$\,\tablefootmark{b} & L$'_{\rm CO}$ & M(H$_2$)\tablefootmark{c}  \\
      & [km~s$^{-1}$]           &                         &  [km~s$^{-1}$] & [Jy~km~s$^{-1}$] & [10$^{10}$~K~km~s$^{-1}$~pc$^{2}$] & [10$^{10}$~M$_{\odot}$]  \\
\hline
\hline
$^{12}$CO(2--1) & $({\rm V_{opt}} -61) \pm 15$ & 1.4161 $\pm$ 0.0001 &  254 $\pm$ 33 & 0.52 $\pm$ 0.06 & 1.36 $\pm$ 0.15 & 1.1 $\pm$ 0.1 \\
\hline
\end{tabular}
\label{tab:flux}
\end{center}
\tablefoot{
\tablefoottext{a}{Position with respect to the optical emission, ${\rm V_{opt}} = 4.25 \times 10^{5}$~km~s$^{-1}$ 
($z = 1.4162$, G10).}
\tablefoottext{b}{Derived from Gaussian fit (see Fig.~\ref{fig:map}, \textit{right panel}).}
\tablefoottext{c}{Assuming a CO luminosity to total gas mass conversion factor of 
$\alpha$ = 0.8~M$_{\odot}$~(K~km~s$^{-1}$~pc$^{2}$)$^{-1}$.}
}
\end{table*}

\section{Results \label{sec:res}}
We clearly detected the $^{12}$CO(2--1) emission line in AGN.1317. 
Figure \ref{fig:map} shows the $^{12}$CO(2--1) intensity map (\textit{left panel}) and 
spectrum of AGN.1317 (\textit{right panel}).
The intensity map was obtained by integrating the line over a 
velocity range of $\sim$360~km~s$^{-1}$, i.e. covering the line emission at 
$-171 < {\rm v} < 193$~km~s$^{-1}$.
The detection is at $\sim$8$\sigma$, but the spatial resolution 
does not allow the galaxy CO emission to be resolved, which is therefore point-like in our observations.

Table~\ref{tab:flux} gives the main parameters derived from our $^{12}$CO(2--1) detection in AGN.1317.
The Gaussian fit applied to the $^{12}$CO(2--1) spectrum (\textit{right panel} of Fig.~\ref{fig:map}, blue curve) 
shows that the molecular gas emission is blue-shifted by $\sim$61~km~s$^{-1}$ with respect 
to the optical emission (${\rm v} = 0$~km~s$^{-1}$ at $z = 1.4162$, G10).
Observations from IRAM PdBI led to a redshift determination of $z = 1.4161 \pm 0.0001$ 
consistent with $z = 1.4162 \pm 0.0005$ from G10 and almost consistent with
$z = 1.4168 \pm 0.0005$ from M12 (see Tables~\ref{table1} and \ref{tab:flux}).  
The $^{12}$CO(2--1) line width is of $\sim$250~km~s$^{-1}$,  in agreement both with 
those typical for field galaxies at redshifts similar to AGN.1317 
\citep[e.g.][]{daddi10,tacconi10} and for intermediate redshift cluster galaxies 
\citep[e.g.][]{geach09,geach11,jablonka13}.

We calculated L$'_{\rm CO}$ (= L$'_{\rm CO(1-0)}$) from the $^{12}$CO(2--1) integrated line emission according to 
\citet{solomon05},
L$'_{\rm CO}$ = 3.25 $\times$ 10$^{7}$ S$_{\rm CO}$$\Delta$${\rm v}$ $\nu_{\rm obs}$$^{-2}$ (1 + $z$)$^{-3}$ D$^2_L$ in K~km~s$^{-1}$~pc$^{2}$, 
where S$_{\rm CO}$$\Delta$${\rm v}$ is the total velocity integrated line flux in Jy~km~s$^{-1}$,
$\nu_{\rm obs}$ the observed frequency in GHz, $z$ the redshift of the source, and 
D$_L$ the luminosity distance in Mpc.
To compute L$'_{\rm CO}$, we assumed a ratio of 1 between the $^{12}$CO(2--1) and $^{12}$CO(1--0) 
luminosities, as expected for a thermalized optically thick CO emission.
Based on these assumptions, AGN.1317 has L$'_{\rm CO} \sim10^{10}$~K~km~s$^{-1}$~pc$^{2}$,
compatible with CO luminosities of $z \gtrsim 1$ galaxies both in cluster \citep[][]{wagg12a} and in field  
\citep[e.g.][]{daddi10,tacconi10}.

It is well established that the CO luminosity linearly traces the molecular gas mass, 
M(H$_2$) = $\alpha$L$'_{\rm CO}$, where M(H$_2$) is defined to include the mass of He so that
M(H$_2$) = M$_{\rm gas}$ \citep[e.g.][]{young91,solomon05}. 
The conversion factor $\alpha$ varies from 
$\alpha$ = 4.6~M$_{\odot}$~(K~km~s$^{-1}$~pc$^{2}$)$^{-1}$ for our Galaxy and normal
spirals to $\alpha$ = 0.8~M$_{\odot}$~(K~km~s$^{-1}$~pc$^{2}$)$^{-1}$ for ultra-luminous 
infrared galaxies (ULIRGs, L$_{\rm FIR} > 10^{12}$~L$_{\odot}$).
Since L$_{\rm FIR}$ of AGN.1317 is unknown, we adopted 
the more conservative ULIRG conversion when estimating H$_2$ mass.
We obtained a value of M(H$_2$) $\sim$ 10$^{10}$~M$_{\odot}$.
We notice that the  $\alpha$ conversion factor depends on various terms, such as the excitation temperature
of the CO, the density of the gas, the gas metallicity, the cosmic ray density,
and the ultraviolet radiation field \citep[e.g.][]{maloney88,boselli02,magrini11}. 
This dependence inevitably introduces a degree 
of uncertainty in the H$_2$ mass estimation and therefore in all parameters involving the gas mass. 
The assumption of a Galactic conversion would increase our mass estimate  
by a factor $\sim$6.
In this sense, treating AGN.1317 as an ULIRG in terms of H$_2$ mass, makes our H$_2$ mass estimation a lower limit.
Despite our very conservative assumption regarding the CO-to-H$_2$
conversion factor, we found that AGN.1317 has a substantial reservoir of
molecular gas. 
This is consistent with what has been observed in the one and only CO detection from a $z \sim 1$ AGN in a cluster \citep{wagg12a}.  
Similar gas contents are also found in high redshift massive objects, such as the bright, field submillimeter galaxies (SMGs) at 
$z \sim 1-3.5$ \citep[][]{greve05} and the colour-selected star-forming galaxies at $z \sim 1.5 - 3$
\citep[see][for a review]{carilli13}.

\section{The origin and fate of molecular gas in AGN.1317 \label{sec:dis}}
As shown in the previous section, our IRAM PdBI observations led to the finding that AGN.1317 
possesses a substantial molecular gas reservoir.
Using information from our new dataset  and from multi-band observations of the overdensity, which 
were already available in the literature, we are able to make hypotheses on the origin of the gas in AGN.1317. 
In a dense environment, such as a galaxy cluster, a non-negligible amount of gas evokes a merging scenario
where a gas fraction might originate from satellites of AGN.1317 and/or 
from the inter-galactic medium.
However, optical images of the cluster show that AGN.1317 is apparently quite isolated within Group1 
(Fig.~\ref{fig:image-cluster}) and, therefore, without nearby companions that survived and from which it could have stripped gas.  
Moreover, the $^{12}$CO(2--1) line width of $\sim$250~km~s$^{-1}$ in AGN.1317 is 
narrow if compared with the broad CO profiles in high-$z$ merging systems, with
line-widths up to 1000 -- 1500~km~s$^{-1}$ 
\citep[e.g.][]{debreuck05,salome12,emonts13,feruglio13}.  
Therefore, we can reasonably exclude recent episodes of  merging to be the  cause of the quantity of gas in AGN.1317. 
The substantial amount of gas is more likely an intrinsic characteristic of the AGN due to  the early phase of  its evolution, when most of its 
baryonic mass was in gas phase.  

Using the spectroscopic NIR information, we are able to derive the current SFR in AGN.1317. 
The narrow and broad H$\alpha$ fluxes are both available in M12. 
As a first order approximation, we have assumed  that most of the narrow H$\alpha$ flux  is originated by SF, 
giving thus a  probe of the young (lifetimes $<20$~Myr), 
massive (M~$>10$~M$_{\odot}$) stellar population,  instead of the AGN activity. 
The use of the narrow component is also supported by the location of the AGN flux ratios in the 
diagnostic Baldwin-Phillips-Terlevich (BPT) diagrams \citep{baldwin81}.
Using the original spectra of AGN.1317 from M12, we measured the narrow component 
of {\sc [O~iii]}5007 and H$\beta$, not available in Table 2 of M12, obtaining log({\sc [O~iii]}5007/H$\beta$) $\sim 0.11$.
With the fluxes of the narrow H$\alpha$,  {\sc [N~ii]}6584, and  {\sc [S~ii]}6716,6731, we obtained 
log({\sc [N~ii]}6584/H$\alpha$) $\sim -0.48$ and log({\sc [S~ii]}6716,6731/H$\alpha$) $\sim -0.63$.
These flux ratios locate AGN.1317 in the starburst region of BPT diagrams from \citet{wang08}.   
We corrected the H$\alpha$ flux for extinction 
using c$_{\beta} = 0.58$ calculated from the ratio of H$\alpha$ and H$\beta$ given in M12. 
Since H$\beta$ flux is not calibrated in M12,  we re-normalized the continuum of the $J$-band spectrum 
to the value of the continuum in the $H$-band spectrum, obtaining a narrow component of the H$\beta$ line  
of $\sim$$6 \times 10^{-17}$~erg~cm$^{-2}$~s$^{-1}$.    
We have therefore estimated the nearly instantaneous SFR in AGN.1317 
using the calibration of \citet{kennicutt98}, 
SFR[M$_{\odot}$~yr$^{-1}$] = 7.9 $\times$ 10$^{-42}$~L(H${\alpha}$) [erg~s$^{-1}$]. 
We obtained a SFR $\sim$65~M$_{\odot}$~yr$^{-1}$, comparable with the most massive galaxies in the overdensity (M12).
This relatively high SFR value indicates that AGN.1317 is actively forming stars and is therefore consuming 
its reservoir of molecular gas.

Assuming that all  H$_2$ is available to sustain the SFR and that the SFR
continues at the current rate, we derive a minimum mass depletion 
time-scale of t$_{\rm depl}$ = M$_{\rm H2}$/SFR $\approx$ 0.2~Gyr
comparable to the  values ($\sim$0.5~Gyr) obtained by \citet{daddi10} for very massive $z = 1.5$ disc galaxies.
This  t$_{\rm depl}$ is obtained treating AGN.1317 as an ULIRG, while
assuming the Galactic $\alpha$ value in the gas mass computation t$_{\rm depl}$ is of
$\approx$1.0~Gyr. 

Summarising, AGN.1317 i) exhibits a large quantity of molecular gas; ii) has a current SFR comparable with the most 
massive galaxies in the overdensity; and iii) maintaining the current SFR, it will consume the molecular gas in about 0.2-1.0~Gyr.
 

\section{Conclusions \label{sec:con}}
We have presented a new IRAM PdBI CO detection in AGN.1317, a source belonging 
to a galaxy cluster at $z \sim 1.4$ and located nearby the centre of the cluster.
Such CO detections in high-$z$ clusters are rare.
From the $^{12}$CO(2--1) line luminosity, we measured  H$_{2}$ mass of 1.1~$\times$10$^{10}$~M$_{\odot}$,
comparable to that in massive SMGs.
The optical images of the cluster show that AGN.1317
is an isolated object, and thus its gas mass might be an intrinsic characteristic associated
with its early evolutionary phase, when most of its baryonic mass was in gas phase, than due
to a merger episode with nearby companions.
The inferred gas mass assumes the CO-to-H$_{2}$ conversion factor adapted for ULIRGs;
if we adopt the Galactic conversion factor, this estimate would increase by a factor of $\sim$6.
The H${\alpha}$-derived SFR is $\sim$65~M$_{\odot}$~yr$^{-1}$, and so AGN.1317 would exhaust its reservoir 
of cold gas in $\sim$0.2-1.0~Gyr.
This relatively high molecular gas content and SFR for an AGN near the centre of a young cluster is compatible with 
models of evolution of galaxy clusters that predict high SFR at the epoch of their formation, 
i.e. $z \sim 1.5$.
Following the predictions of these models,  at $z \sim 1$, the SFR would  rapidly be quenched  through environmental 
effects starting from the innermost regions of clusters. 

\begin{acknowledgements}
We thank the anonymous referee for useful comments and suggestions 
which improved the quality of the manuscript.
We also thank the IRAM PdBI staff for help provided during observations and data reduction.
The research leading to these results has received funding from the European 
Commission Seventh Framework Programme (FP/2007-2013) under grant agreement No 283393 (RadioNet3).
This research has made use of the NASA/IPAC Extragalactic Database (NED).
We thank A. Galametz for making available the $B$-band image of  the field around the radio galaxy 7C 1756+6520.  
\end{acknowledgements}

\end{document}